# Fenomenologia da Supercondutividade e Supercondutores Mesoscópicos

*Phenomenology of the superconductivity and mesoscopic superconductors*

*Rafael Zadorosny[1], Alice Presotto[1], Elwis C. S. Duarte[1], Edson Sardella[2]*

[1] Departamento de Física e Química, Univ Estadual Paulista – Unesp, CP 31, CEP 15385-000, Ilha Solteira, SP

[2] Departamento de Física, Faculdade de Ciências, Univ Estadual Paulista – Unesp, CP 473, CEP 17033-360, Bauru, SP

**Resumo**

Em geral, quando a teoria fenomenológica de Ginzburg-Landau para supercondutores é trabalhada, pouco se esclarece aos alunos sobre a origem da mesma. Esta tem como base conceitos termodinâmicos como fenômenos críticos e transições de fase que, se devidamente tratados, enriquecem de sobremaneira a aula sobre tal assunto. Assim, neste trabalho apresentamos uma sequência para o desenvolvimento da principal teoria fenomenológica da supercondutividade. Iniciamos com uma breve introdução aos fenômenos críticos e transições de fase e, então, desenvolvemos a teoria de Landau para transições de fase de segunda ordem. Após isso, expomos a teoria de Ginzburg-Landau, e a teoria de Ginzburg-Landau dependente do tempo. Aplicamos esta última para o caso de dois sistemas supercondutores mesoscópicos, um homogêneo e outro com defeitos superficiais. Observa-se que os defeitos interferem de sobremaneira na dinâmica de vórtices do sistema.

***Palavras chave:*** *Ginzburg-Landau, mesoscópico, vórtice, supercondutor*

**Abstract**

Usually, during the classes, the origin of the phenomenological theory proposed by Ginzburg-Landau is not treated. The fundamentals of such theory are based on thermodynamics concepts like criticality and phase transitions that, if used, could enrich a lot the classes. Thus, in this work we present a sequence for the development of this major phenomenological theory for superconductors. We will begin with a brief introduction to the criticality and phase transitions phenomena e then we will describe the Landau theory for second order phase transitions. After that, we will introduce the Ginzburg-Landau theory and the time-dependent Ginzburg-Landau theory. As an application of this last one, we will study two mesoscopic systems. One of them a homogeneous sample and the other one a sample with surface defects. We show that the defects strongly influences the vortex dynamics of the system.

## 1. Introdução

Há mais de cem anos os fenômenos críticos são observados. Esses fenômenos são assim denominados por ocorrerem próximos a um ponto crítico que, por sua vez, pode ser entendido como o ponto do diagrama de fase de um sistema onde a transição de fase é de segunda ordem. Nas suas proximidades, diferentes sistemas comportam-se de modo semelhante, obedecendo a leis de potências cujos expoentes não são inteiros, os denominados expoentes críticos [1, 2].

Várias teorias foram propostas na tentativa de explicar esses fenômenos físicos. Destacamos nesse trabalho a simples, porém elegante teoria para transições de fase de segunda ordem proposta por Landau em 1937 [3]. Tal teoria é baseada na expansão da energia livre do sistema estudado em função do seu parâmetro de ordem. O parâmetro de ordem, por sua vez, é uma quantidade definida como zero em uma fase do sistema (fase desordenada, acima da temperatura crítica) e com valores diferentes de zero em outra (fase ordenada, abaixo da temperatura crítica).

A teoria de Landau nos dá elementos importantes para o desenvolvimento da teoria de Ginzburg-Landau (GL) [4-7] publicada em 1950 a qual descreve satisfatoriamente vários fenômenos relacionados à supercondutividade. Por seu aspecto fenomenológico e simples, os pesquisadores da área receberam, à época de sua publicação, tal trabalho com ceticismo mas, em 1959, quando Gorkov [4] mostrou que as equações GL eram um caso particular da teoria supercondutora de primeiros princípios BCS [8,9], esta se popularizou no meio. Em 1966, Schimid [10] inseriu uma dependência temporal às equações GL e, dessa forma, a dinâmica de sistemas fora do equilíbrio foi possível de ser estudada.

## 2. Breve abordagem sobre Fenômenos Críticos e Transições de Fase

Todos os fenômenos que ocorrem nas proximidades de um ponto crítico são denominados Fenômenos Críticos (FC). Este ponto, em geral, é o final de uma curva finita que separa duas fases distintas de um sistema, como por exemplo a curva de vaporização da água, que separa o estado líquido do gasoso. Geralmente denotamos por $T_C$ e $P_C$, por exemplo, a temperatura e pressão do ponto crítico. No contorno deste ponto, é possível passar de um estado a outro da matéria (de vapor para líquido, por exemplo) sem haver uma mudança abrupta entre os estados [1]. Este comportamento pode ser descrito por um parâmetro de ordem característico de cada substância. Outro fato interessante é que, embora diferentes sistemas apresentem diferentes fenômenos de transição de fase, na região próxima ao ponto crítico todos têm o mesmo comportamento.

Assim, é possível identificar a ordem das transições de fase pelo comportamento do parâmetro de ordem do sistema. Se esse mudar de forma contínua entre uma fase e outra, a transição é de segunda ordem, mas se for descontínuo, a transição será de primeira ordem. Essa definição é diferente da inserida por Ehrenfest, onde a ordem da transição está relacionada com a ordem da derivada do potencial termodinâmico adequado. Se a primeira derivada for descontínua, a transição será de primeira ordem, mas se a segunda derivada for aquela que apresentar alguma descontinuidade, mantendo-se a primeira derivada contínua, a transição será de segunda ordem e assim por diante. Nos casos analisados nesse trabalho, ambas as definições serão satisfeitas.

Com esta breve noção sobre fenômenos críticos, na próxima seção será analisada a Teoria de Landau para transições de fase de segunda ordem. Por essa teoria, a energia livre do sistema é descrita em termos do parâmetro de ordem nas proximidades do ponto crítico.

## 3. A Teoria de Landau para transições de segunda ordem

Essa teoria tem por base a expansão da energia livre do sistema (esta pode ser entendida como a energia que este possui e possível de ser transformada em trabalho) em potências do parâmetro de ordem, $\eta$, específico de cada sistema. Dessa forma, a continuidade da mudança de estado na transição de fase de segunda ordem é expressa matematicamente pelo fato de $\eta$ ter valores pequenos perto do ponto crítico. Assim, considerando a vizinhança desse ponto, expandimos a energia livre de Gibbs, por exemplo, em potências de $\eta$

$$G = G_0 + \alpha\eta + a\eta^2 + c\eta^3 + b\eta^4 + ... \qquad (1)$$

onde a expansão foi feita até quarta ordem devido ao baixo valor de $\eta$ e os coeficientes, $\alpha$, $a$, $c$ e $b$ são funções da pressão (P) e da temperatura (T).

Para obtermos $\eta$ devemos minimizar a energia livre:

$$\frac{\partial G}{\partial \eta} = \alpha + 2a\eta + 3c\eta^2 + 4b\eta^3 = 0 \quad (2)$$

$$\alpha + \eta(2a + 3c\eta + 4b\eta^2) = 0 \quad (3)$$

Com isso, as possíveis soluções para a equação (3) são:

$$\begin{aligned} \alpha = 0 \quad e \quad \eta = 0 \\ \alpha = 0 \quad e \quad \eta \neq 0 \end{aligned} \quad (4)$$

De todas as soluções possíveis, vemos que $\alpha$ é sempre zero, independente do valor de $\eta$.

Para que a condição de mínimo seja satisfeita, a derivada segunda da energia livre deve ser positiva, com isso temos:

$$\frac{\partial^2 G}{\partial \eta^2} = 2a + 6c\eta + 12b\eta^2 \quad (5)$$

Para $\eta = 0$, que corresponde a $T > T_C$, devido à definição do parâmetro de ordem, temos que a equação (5) só será positiva se $a$ for positivo.

Para $\eta \neq 0$ (ou seja, $T < T_C$), vamos analisar a segunda derivada $G$ desprezando a constante $b$ na equação (3). Obtemos, então, uma expressão para $\eta$ dada pela equação (6).

$$\eta = -\frac{2a}{3c} \quad (6)$$

Substituindo a equação (6) na equação (5) sem a constante $b$ temos:

$$\frac{\partial^2 G}{\partial \eta^2} = -4a \quad (7)$$

então, para que a equação (7) seja positiva, a constante $a$ deve ser negativa. Assim, já que $a$ é positiva acima de $T_C$ e negativa abaixo dessa temperatura, concluímos que no ponto crítico, ou seja, em $T_C$, a constante $a$ é nula.

Nessa teoria assume-se que a função $a(P,T)$ não apresenta singularidades no ponto crítico, então Landau propôs a seguinte expressão:

$$a = a_0(T - T_C) \quad (8)$$

onde:

$$\begin{aligned} a > 0 \quad para \quad T > T_C \\ a = 0 \quad para \quad T = T_C \\ a < 0 \quad para \quad T < T_C \end{aligned} \quad (9)$$

Reescrevendo a equação (1) e a sua segunda derivada já com $\alpha = 0$ e aplicando-as no ponto crítico temos:

$$G_C = G_0 + c_C\eta^3 + b_C\eta^4 \quad (10)$$

$$\frac{\partial^2 G_C}{\partial \eta^2} = 6c_C\eta + 12b_C\eta^2 \quad (11)$$

Da equação (11) obtemos que se $c_c = 0$, $b_c > 0$. Dessa forma, se $b > 0$ no ponto crítico, ele será positivo para qualquer ponto nas proximidades dessa região.

Resta-nos duas soluções para $c$. Para $c = 0$, significa que há uma linha de pontos críticos no plano $PT$. Porém, se $c \neq 0$, a transição de segunda ordem só ocorrerá em pontos isolados, como em uma transição líquido-gás.

Escolhemos $c = 0$ [3] pois este é o caso mais interessante, onde há uma linha de pontos críticos, o qual é válido para os supercondutores também. Assim a equação (1) adquire a forma:

$$G = G_0 + a_0(T - T_C)\eta^2 + b\eta^4 \qquad (12)$$

A Figura 1 mostra duas curvas da equação (12) para valores positivos e negativos do coeficiente *a*. Note que a curva apresenta mínimos locais fora da origem do sistema de coordenadas para *a*<0. Landau também considerou o coeficiente *b* como sendo praticamente constante quando $T \to T_C$.

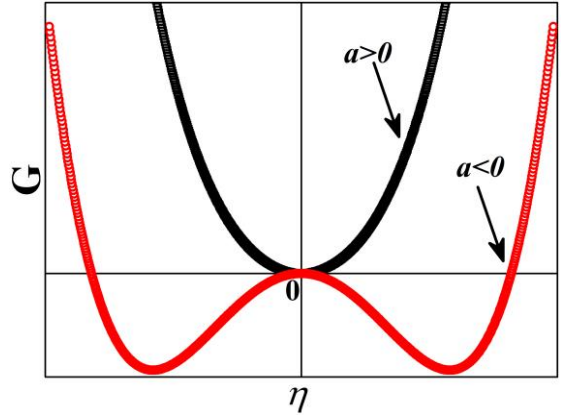

**Figura 1:** Energia livre de Gibbs em função do parâmetro de ordem *η* para os valores de *a* da equação (9)

Agora podemos obter uma expressão para *η* dependente da temperatura próxima ao ponto crítico minimizando a energia livre de Gibbs em relação ao próprio *η*. Com isso temos:

$$\eta^2 = -\frac{a_0(T - T_C)}{2b} = \frac{|a|}{2b} \qquad (13)$$

Desprezando as potências mais altas de *η*, encontramos a entropia do sistema.

$$S = -\frac{\partial G}{\partial T} = S_0 - a_0\eta^2$$

Com isso temos:

$$\begin{aligned} S &= S_0 & para \quad T > T_C \\ S &= S_0 + \frac{a_0^2}{2b}(T - T_C) & para \quad T < T_C \end{aligned} \qquad (14)$$

No ponto crítico a entropia fica $S = S_0$, ou seja, ela é contínua nesse ponto, como vemos na Figura 2 (a).

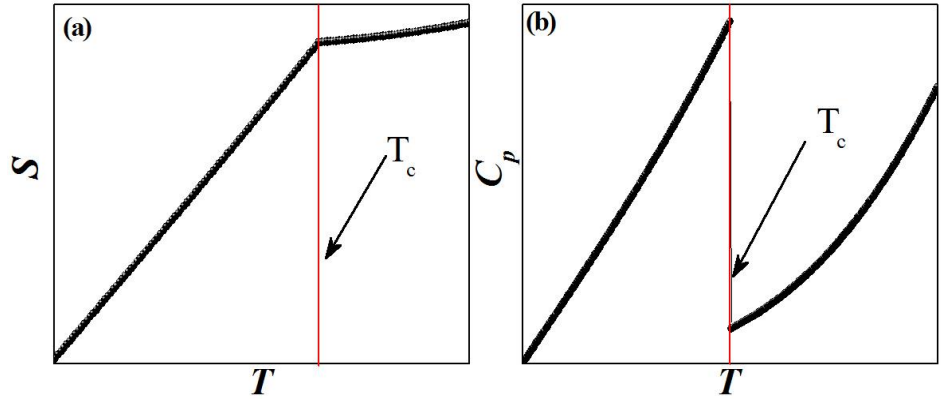

**Figura 2:** Continuidade da entropia na transição de fase e descontinuidade do calor específico para a mesma transição. Esse comportamento é típico de uma transição de segunda ordem.

O calor específico é dado por:

$$C_P = T\frac{\partial S}{\partial T}\bigg|_P = -T\frac{\partial^2 G}{\partial T^2}\bigg|_P$$

Assim, das equações (14) temos:

$$\begin{aligned} C_P &= C_{P0} & \text{para} \quad T > T_C \\ C_P &= C_{P0} + T\frac{a_0^2}{2b} & \text{para} \quad T < T_C \end{aligned} \quad (15)$$

Na Figura 2 (b), é mostrada a descontinuidade do calor específico no ponto crítico.

### 4. A teoria de Ginzburg – Landau (GL) para supercondutores

A teoria fenomenológica de Ginzburg-Landau (GL) segue a mesma metodologia da teoria de Landau para transições de fase de segunda ordem já que, na ausência de campo magnético, a transição de fase normal-supercondutor também é de segunda ordem. Essa teoria foi publicada em 1950 e uma das motivações de tal trabalho foi a necessidade de descrever a destruição da supercondutividade por um campo magnético ou uma corrente elétrica. Maiores detalhes sobre a teoria GL podem ser obtidos na Ref. 7.

Na teoria GL o parâmetro de ordem do supercondutor é uma função complexa $\Psi(\vec{r})$ que é interpretada como a função de onda "efetiva" dos superelétrons (portadores de carga do estado supercondutor) e $|\Psi(\vec{r})|^2 = n_s$, onde $n_s$ é a densidade dos superelétrons.

$$\Psi(\vec{r}) = |\Psi(\vec{r})|e^{i\phi} = \sqrt{n_s}\,e^{i\phi} \quad (16)$$

Seguindo as mesmas considerações estabelecidas por Landau e analisadas na seção 3, escrevemos a energia livre de Gibbs do supercondutor na ausência de campo magnético, equação 17. É interessante frisar que alguns autores escrevem a energia do estado supercondutor em termos da energia livre de Helmholtz. Porém, esta não leva em consideração o trabalho útil realizado pela magnetização, assim, escreveremos a energia deste estado pela energia livre de Gibbs. Contudo, ao minimiza-las, nota-se que as equações GL independem da energia livre escolhida.

$$G_S = G_n + \int\left\{\alpha|\Psi|^2 + \frac{\beta}{2}|\Psi|^4 + \frac{1}{4m}\left|-i\hbar\nabla\Psi\right|^2\right\}d^3r \quad (17)$$

Onde $G_n$ é a energia livre do estado normal, $\hbar$ é a constante de Planck, $\beta$ é um coeficiente positivo independente da temperatura (é a constante $b$ da teoria de Landau multiplicada por dois), $\alpha$ é função da temperatura dada pela equação (8):

$$\alpha = \alpha_0 (T - T_C) \tag{18}$$

sendo $T_C$ a temperatura crítica acima da qual o material vai para o estado normal e o coeficiente $\frac{1}{4m} \left| -i\hbar \nabla \Psi \right|^2$ é a densidade de energia cinética dos superelétrons inserida na expansão da energia livre pelo fato do parâmetro de ordem do supercondutor ser dependente da posição.

Na presença de campo magnético, a equação (17) deve ser corrigida por dois termos:

- A densidade de energia magnética, dada por $\frac{B^2}{8\pi} - \vec{H} \bullet \vec{M}$, onde $\vec{B}$ é o campo de indução magnética que no interior do supercondutor é zero e assume o valor do campo magnético aplicado, $\vec{H}$, no estado normal e, por fim, $\vec{M}$ é a magnetização do material supercondutor, ou seja, $\vec{M} = 0$ para $T > T_C$ (estado normal) e $\vec{M} = -\frac{\vec{H}}{4\pi}$ para $T < T_C$. Note que estamos trabalhando no sistema cgs.

- A substituição do operador $-i\hbar \nabla$ por $-i\hbar \nabla - \frac{2e}{c} \vec{A}$, onde $2e$ é a carga do superelétron, $\vec{A}$ é o potencial vetor e $c$ a velocidade da luz.

Com isso a equação (17) adquire a forma:

$$G_S = G_n + \int \left\{ \alpha |\Psi|^2 + \frac{\beta}{2} |\Psi|^4 + \frac{1}{4m} \left| -i\hbar \nabla \Psi - \frac{2e}{c} \vec{A} \Psi \right|^2 + \frac{B^2}{8\pi} - \vec{H} \bullet \vec{M} \right\} d^3r \tag{19}$$

Minimizando $G_s$ com relação a $\delta \Psi^*$ obtemos a primeira equação GL e sua condição de contorno:

$$\alpha \Psi + \beta \Psi |\Psi|^2 + \frac{1}{4m} \left( -i\hbar \nabla - \frac{2e}{c} \vec{A} \right)^2 \Psi = 0 \tag{20}$$

$$\left( -i\hbar \nabla \Psi - \frac{2e}{c} \vec{A} \Psi \right) \bullet \vec{n} = 0 \tag{21}$$

onde $\vec{n}$ é o vetor normal à superfície do supercondutor.

A equação (21) nos diz que não há supercorrentes passando pela interface normal – supercondutor. A minimização de $G_S$ com respeito a $\Psi$ dá o complexo conjugado da equação (20). Resta-nos, então, minimizar $G_S$ em relação ao potencial vetor $\vec{A}$. Assim, obtemos a segunda equação de GL.

$$\nabla \times \nabla \times \vec{A} = -\frac{2\pi i \hbar e}{mc} \left( \Psi^* \nabla \Psi - \Psi \nabla \Psi^* \right) - \frac{8\pi e^2}{mc^2} \vec{A} |\Psi|^2 \tag{22}$$

Esta equação ainda pode ser reescrita usando a lei de Ampère, $\nabla \times \vec{B} = \nabla \times \nabla \times \vec{A} = \frac{4\pi}{c} \vec{J}_S$, adquirindo a forma:

$$\vec{J}_S = -\frac{i\hbar e}{2m} \left( \Psi^* \nabla \Psi - \Psi \nabla \Psi^* \right) - \frac{2e^2}{mc} \vec{A} |\Psi|^2 \tag{23}$$

Embora as equações GL sejam provenientes de uma teoria fenomenológica simples, elas nos permitem descrever uma gama de propriedades relacionadas aos supercondutores. Não demonstraremos neste trabalho mas, utilizando algumas considerações e aplicando-as nas equações GL, obtém-se a quantização do campo magnético e os comprimentos característicos dos supercondutores, ou seja, a profundidade de penetração de London, $\lambda$, e o comprimento de coerência, $\xi$.

O $\lambda$ é devido ao fato de que, embora o supercondutor exclua o campo magnético de seu interior, as correntes que o blindam se distribuem por uma certa região de espessura $\lambda$. Consequentemente, o fluxo magnético não cai abruptamente para zero dentro do supercondutor, mas sim ele o penetra decaindo ao longo da distribuição das correntes de blindagem. Dessa forma, $\lambda$ é conhecido por profundidade de penetração e nos indica o quanto o fluxo magnético (ou, o campo magnético) penetra no supercondutor.

Já o conceito de $\xi$ foi formulado em 1953 por Pippard [11,12]. Ele considerou que a densidade de superelétrons não poderia variar rapidamente com a posição, mas variaria apreciavelmente dentro de uma certa distância. Uma conseqüência da existência de $\xi$ é que o contorno entre uma região normal e uma região supercondutora deve ter uma largura finita (i.e., não nula), pois a densidade de superelétrons varia de zero na primeira região até um valor máximo $n_s$ dentro do supercondutor sendo, então, $\xi$, uma medida da espessura dessa região de variação. A razão $\lambda/\xi$ é conhecida por parâmetro de GL, $\kappa$, e, em geral, independe de T. Este parâmetro é muito importante no cálculo da densidade de energia superficial, $\sigma$. Esta, contudo, se anula para $\kappa = 1/\sqrt{2}$ e pode assumir valores positivos e negativos. Com isso, os supercondutores são divididos em duas classes: os do tipo I (SCI), para $\kappa < 1/\sqrt{2}$, onde $\sigma > 0$, e os do tipo II (SCII), para $\kappa > 1/\sqrt{2}$, onde $\sigma < 0$. Neste último caso, torna-se energeticamente favorável a penetração de fluxo quantizado ($\Phi_0$) no interior dos SCII. Estes *fluxóides* possuem um núcleo de material normal e supercorrentes fluem ao seu redor para confinar o fluxo em tal região que, por sua vez, possui raio igual a $\xi$. Assim, ao conjunto núcleo normal mais supercorrentes denominamos por vórtice (ver inset do painel inferior da Figura 3). Nos painéis principais da Figura 3, são mostrados os comportamentos magnéticos típicos dos SCI (superior) e SCII (inferior). Neste último, o estado misto está caracterizado por uma exemplificação de rede triangular de vórtices, ou seja, a rede de Abrikosov, como será descrito no próximo parágrafo. Os insets do painel superior exemplificam as curvas de $\lambda$ e $\xi$ para as situações onde $\kappa < 1/\sqrt{2}$ (painel (a)) e $\kappa > 1/\sqrt{2}$ (painel (b)).

Em 1952, dois anos após a publicação da teoria GL, Alexei Abrikosov, começou a estudar o caso de supercondutores com $\kappa > 1/\sqrt{2}$. Nesse estudo, Abrikosov descobriu que para essa classe de supercondutores, acima de certo campo crítico inferior, $H_{C1}$, se torna energeticamente favorável a presença de vórtices no interior dos materiais. Além disso, ele previu que conforme o campo externo aumenta sua intensidade, os vórtices se dispõem em uma rede triangular, hoje conhecida por rede de Abrikosov (ver exemplificação no painel inferior da Figura 3). À medida que o campo externo aumenta, os vórtices se tornam mais próximos uns dos outros e começam a se sobrepor. No campo denominado $H_{C2}$, os núcleos dos vórtices estão totalmente sobrepostos e o estado supercondutor é destruído.

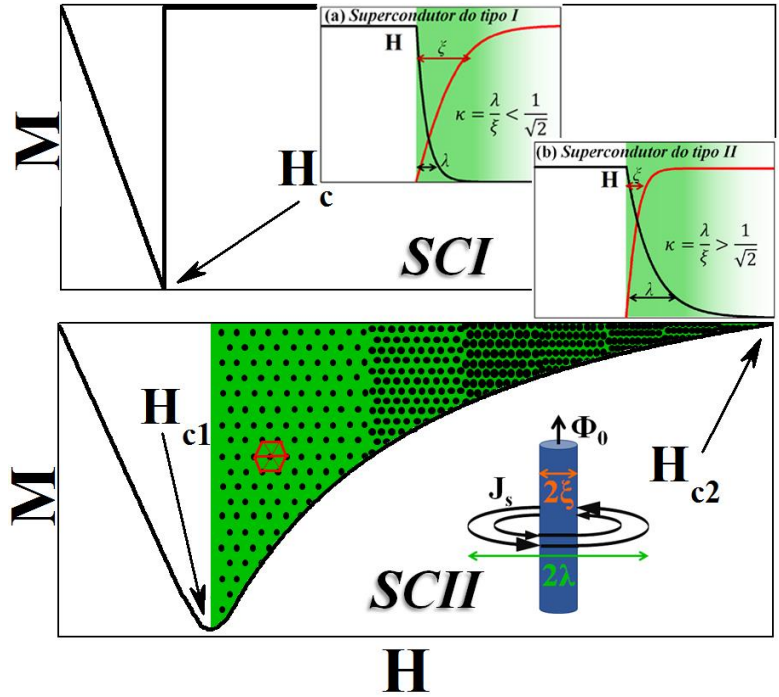

**Figura 3:** Painel superior exemplifica a resposta magnética de um SCI por uma curva de magnetização em função do campo magnético externo aplicado. No painel inferior, é exemplificada a resposta de um SCII onde o estado misto é caracterizado pela presença de vórtices que se arranjam formando a rede de Abrikosov. Os insets mostram as curvas típicas da penetração do campo magnético (medida por λ) e da variação do parâmetro de ordem (medido por ξ) para os SCI (a) e SCII(b).

Essas previsões ficaram "engavetadas" até 1957 quando Abrikosov finalmente as publicou. Em 2003 ele foi laureado com o prêmio Nobel de Física, que dividiu com Vitaly Ginzburg e Anthony Leggett – o primeiro é o G da teoria GL, o segundo trabalhou com condensação de Bose-Einstein, trabalho que motivou Landau, orientador de Abrikosov, a reunir o seu grupo para dizer que Feynmam, um dos pesquisadores dessa área, havia resolvido o problema de superfluidez. Foi exatamente esse fato que despertou Abrikosov e incentivou sua publicação [10, 11].

### 5. A teoria de Ginzburg – Landau dependente do tempo (TDGL) e supercondutores mesoscópicos

Em 1966, Schmid publicou um trabalho onde atribuída uma taxa de variação temporal ao parâmetro de ordem e ao potencial vetor na teoria GL. As equações resultantes ficaram conhecidas por TDGL (time-dependent Ginzburg-Landau) [13]. Dessa forma, há a possibilidade de acompanhar a evolução dos vórtices em estados metaestáveis, ou seja, estados entre dois estados estacionários consecutivos. Schmid também fez a inclusão do potencial escalar elétrico que descreve um sistema supercondutor na presença de uma corrente de transporte.

A 1ª equação TDGL é

$$\frac{\hbar^2}{2m_s D}\left(\frac{\partial}{\partial t} + i\hbar e_s \Phi\right)\Psi = -\frac{\delta G_s}{\delta \Psi} \qquad (24)$$

e a 2ª equação TDGL

$$\beta\left(\nabla\Phi + \frac{\partial \vec{A}}{\partial t}\right)\Psi = -\frac{\delta G_s}{\delta \vec{A}} \qquad (25)$$

onde, $\Phi$ é o potencial escalar, $\beta$ é a condutividade elétrica, D é um coeficiente de difusão, o termo entre parênteses em (25) é o campo elétrico, **E**, de acordo com a lei de Faraday.

O lado direito das equações (24) e (25) correspondem, respectivamente, à 1ª e à 2ª equações GL. A equação (25) também pode ser descrita em termos das densidades de correntes, que fica da seguinte forma:

$$\vec{J} = \beta\vec{E} + \vec{J}_s \qquad (26)$$

Muitas vezes é interessante trabalhar com as equações TDGL em sua forma adimensional. Com isso, as distâncias estão em unidades de $\xi(0)$, os campos em termos de $H_{C2}(0)$, T em unidade de $T_C$ e o tempo em unidades do tempo característico $t_0 = \pi\hbar/8k_B T_C$, com $k_B$ a constante de Boltzmann.

$$\left(\frac{\partial}{\partial t} + i\Phi\right)\Psi = -\left(i\nabla - \vec{A}\right)^2 \Psi + (1-T)\Psi(1-|\Psi|^2) \qquad (27)$$

$$\beta\left(\nabla\Phi + \frac{\partial \vec{A}}{\partial t}\right) = (1-T)\Re\left[\Psi*(-i\nabla - \vec{A})\Psi\right] - \kappa^2 \nabla \times \nabla \times \vec{A} \qquad (28)$$

onde $\vec{B} = \nabla \times \vec{A}$ é o campo a magnético local e, $\beta$ é a condutividade elétrica, $\kappa$ é o parâmetro de Ginzburg-Landau e $\Phi$ é o potencial escalar. A densidade de supercorrente é dada por:

$$\vec{J}_s = (1-T)\Re\left(\Psi*(-i\nabla - \vec{A})\Psi\right) \qquad (29)$$

Esta teoria tem sido muito usada para o estudo de supercondutores mesoscópicos. Tais materiais tem dimensões da ordem de $\lambda$ e, por serem muito pequenos, efeitos de confinamento dominam a dinâmica de vórtices e alteram de sobremaneira os comportamentos dos supercondutores quando comparados a materiais volumétricos, os ditos bulks [16-18]. De uma forma geral, no estado estacionário os vórtices, que nos materiais bulk se arranjam numa rede triangular, a rede de Abrikosov, nos mesoscópicos a influência da superfície é tão considerável que o arranjo dos vórtices passa a seguir a simetria do sistema [19,20]. Outro comportamento exótico é a formação de vórtices gigantes, i.e., vórtices que possuem mais do que um quantum de fluxo magnético em seu núcleo [17,20]. Na Figura 4 são mostradas a variação de $\Psi$ ao longo de dois sistemas mesoscópicos quadrados de tamanhos laterais de $12\xi(0)$ simulados em $T=0,1T_c$. Um deles é uma amostra homogênea, sem qualquer defeito, Figura 4(a) e (c). No outro, Figura 4(b) e (d), foram inseridos cinco defeitos em uma das superfícies do mesmo. Estes defeitos tem largura e profundidade de $0,25\xi(0)$. Na Figura 4(a) e (b) é mostrado um estado não estacionário durante a primeira penetração e, nos respectivos insets, o estado estacionário. Note que no sistema homogêneo, dois vórtices penetram ao mesmo tempo enquanto que no sistema com defeitos, ocorre a penetração de apenas um vórtice. As linhas pretas em tais imagens indicam as correntes de blindagem. Note que estas se deformam ao contornar os defeitos e geram um acúmulo local de linhas de corrente o que, consequentemente, causa um aumento local na corrente crítica do material. Este efeito é conhecido por "current crowding" [21]. Na Figura 4 (c) e (d) são mostrados os arranjos dos vórtices no dois sistemas. Note que, enquanto no sistema homogêneo os vórtices penetrados se arranjam de tal forma a repetir a simetria deste (contorno quadrado), no sistema com defeitos o arranjo é totalmente alterado. Isto mostra que mesmo defeitos muito pequenos podem gerar grandes alterações em sistemas mesoscópicos, tanto na dinâmica de penetração de vórtices quanto no arranjo dos mesmos.

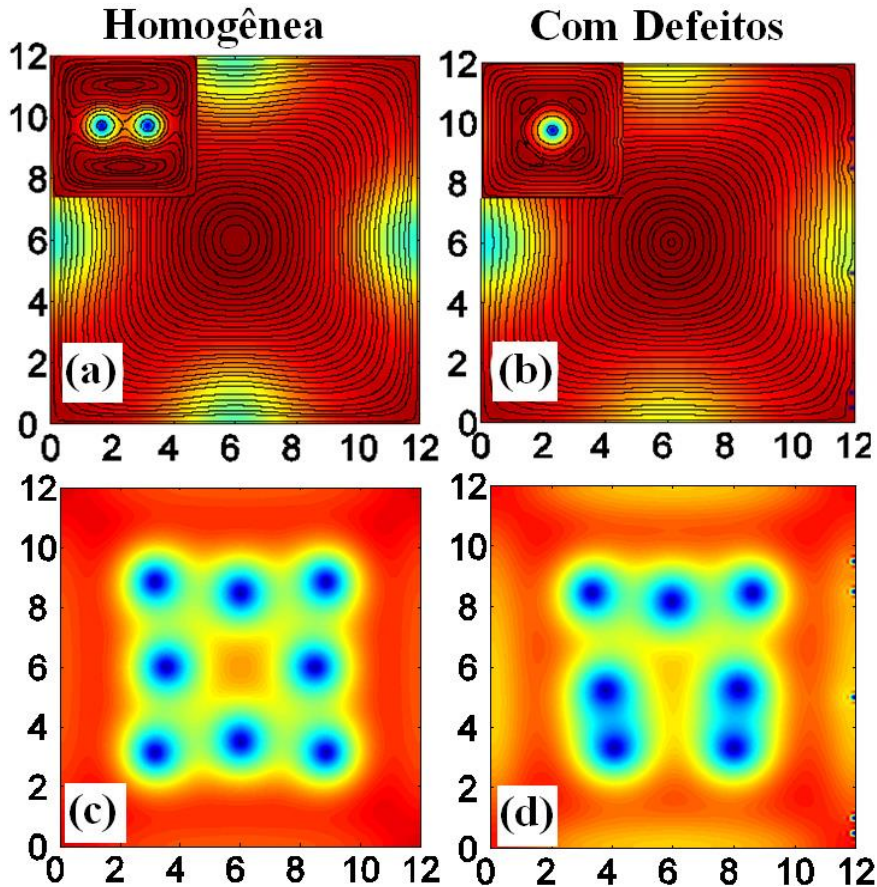

**Figura 4:** Mapeamento de Ψ ao longo de dois sistemas mesoscópicos de tamanho lateral 12ξ(0). Na simulação foi usado κ=5, que equivale à uma liga Pb-In, com λ = 150 nm, ξ = 30 nm e $T_C$=7.0 K [5]. Em (a) é mostrado o estado não estacionário da primeira penetração para uma amostra homogênea e o inset mostra o respectivo estado estacionário. Em (b) o mesmo comportamento é mostrado, porém, para um sistema com cinco defeitos em uma das superfícies da amostra. As linha que contornam tais imagens representam as correntes de blindagem onde, no contorno dos defeitos surgem o efeito de acúmulo das linhas de corrente. Em (c) e (d) são mostrados os arranjos dos vórtices para o sistema homogêneo e com defeitos, respectivamente.

**Conclusão**

Iniciamos nossa análise definindo o conceito de ponto crítico e identificando que, na vizinhança deste, a transição de fase é de segunda ordem. Nesse contexto analisamos a teoria de Landau para tais transições. Esta é baseada na expansão da energia livre do sistema em termos do seu parâmetro de ordem.

Com base na teoria de Landau, introduzimos a teoria fenomenológica de Ginzburg – Landau (GL). Esta é uma importante teoria capaz de explicar muitas características apresentadas pelos supercondutores. Por ela, obtém-se os comprimentos característicos dos supercondutores, i.e., a profundidade de penetração, λ, e o comprimento de coerência, ξ. A razão do primeiro pelo segundo é o parâmetro de Ginzburg-Landau, κ. Assim, para $\kappa < 1/\sqrt{2}$, a energia superficial é positiva, e tem-se os supercondutores do tipo I; por outro lado, para $\kappa > 1/\sqrt{2}$, a energia superficial é negativa, o que define os supercondutores do tipo II. Esses últimos, além de apresentarem exclusão total de fluxo magnético do seu interior até um campo crítico inferior $H_{C1}$, acima deste passa a ser energeticamente favorável a penetração de fluxo quantizado (os vórtices), em seu interior. Esse estado misto, onde há a convivência de estado normal com o estado supercondutor, perdura até um campo crítico superior denominado $H_{C2}$, acima do qual há a transição do estado supercondutor para o estado normal. Em 1966, com a inserção da variação temporal na teoria GL, conhecida como TDGL, foi possível estudar regimes dinâmicos em sistemas supercondutores. Assim, como exemplificação, analisamos dois sistemas mesoscópicos, sendo um deles homogêneo e o outro com defeitos superficiais. Nota-se que, mesmo pequenos, com cerca de 1/4 do raio de um vórtice, os defeitos interferem de sobremaneira na dinâmica de vórtices do sistema.

**Agradecimentos**